\documentclass[preprint,12pt]{revtex4}
\usepackage{bm}
\usepackage{amsmath}
\usepackage{amssymb}
\usepackage{latexsym}
\usepackage{amsfonts}
\usepackage{epsfig}
\usepackage{psfrag}
\usepackage{color}
\usepackage{dsfont}
\usepackage{hyperref}

\makeatletter
\renewcommand{\p@subsection}{}
\renewcommand{\p@subsubsection}{}
\makeatother

\def\new{\newcommand}
\def\renew{\renewcommand}
\new{\eq}[1]{(\ref{#1})}

\new{\bd}{\begin{displaymath}}
\new{\ed}{\end{displaymath}}
\new{\be}{\begin{equation}}
\new{\ee}{\end{equation}}
\new{\bi}{\begin{itemize}}
\new{\ei}{\end{itemize}}
\new{\ba}{\begin{eqnarray}}
\new{\ea}{\end{eqnarray}}
\new{\bo}{\boldmath}
\new{\un}{\unboldmath}
\new{\mx}{\mbox}
\renewcommand{\bold}[1]{\mbox{\boldmath$#1$\unboldmath}}
\new{\bqu}{\begin{quote}}
\new{\equ}{\end{quote}}
\new{\sbs}[2]{\mbox{${#1}_{#2}$}}
\new{\subvel}[3]{\mbox{$#1_{#2},\ldots,#1_{#3}$}}
\new{\subvec}[3]{\mbox{$#1_{#2},\cdots,#1_{#3}$}}
\new{\avec}[2]{\mbox{${#1}_{1},\ldots,{#1}_{#2}$}}
\new{\spin}[2]{ \mbox{$ \bold{\sigma}_{#1} {\bf \cdot \hat{#2}}   $}}
\new{\Sc}{Schr\"{o}dinger}
\new{\sa}{self-adjoint}
\new{\se}{Schr\"odinger's equation}
\new{\BM}{Bohmian mechanics}
\new{\qf}{quantum formalism}
\new{\qt}{quantum theory}
\new{\qm}{quantum mechanics}
\new{\qp}{quantum potential}
\new{\wf}{wave function}
\new{\ewf}{effective wave function}
\new{\cwf}{conditional wave function}
\new{\oqm}{orthodox \qm}
\new{\oi}{orthodox interpretation}
\new{\qe}{quantum equilibrium}
\new{\rv}{random variable}
\new{\hv}{hidden variable}
 \new{\PV}{projection--valued--measure}
\new{\POV}{positive operator valued measure}
\new{\Ae}{Afshar's experiment}
\new{\PC}{`principle of complementarity'}
\new{\Ci}{Copenhagen interpretation}
\new{\born}{\rho=|\psi|^2}
\new{\ai}{\alpha\in {\cal I}}
\renew{\a}{\alpha}
\new{\suma}{\sum_{\a \in I }}
\new{\ot}{\otimes}
\new{\bigo}{\bigoplus}
\new{\la}{\lambda_{\a}}
\new{\biga}{\bigoplus_{\a}}
\new{\psia}{\psi_{\a}}
\new{\Phia}{\Phi_{\a}}
\new{\Ha}{{\H}_{\a}}
\renew{\H}{\mbox{${\cal H}$}}
\new{\R}{\mbox{${\rm I\!R}$}}
\new{\F}{{\cal F }}
\new{\E}{\mbox{${\cal E}$}}
\new{\Ex}{\mbox{${\cal E}$}}
\renew{\P}{\mbox{${\rm I\!P}$}}
\new{\M}{\mbox{${\cal M}$}}
\new{\N}{\mbox{${\cal N}$}}
\new{\Pa}{P_{\a}}
\new{\LA}{\Lambda}
\renew{\O}{{O}}
\new{\As}{{R}}
\new{\Aa}{R_{\a}}
\new{\Aad}{R^{\dagger}_{\a}}
\new{\Al}{R_{\lambda}}
\new{\Ald}{R^{\dagger}_{\lambda}}
\new{\Ay}{R_y}
\new{\Ayd}{R^{\dagger}_{y}}
\new{\pr}{\mbox{Prob}\,}
\new{\prp}{\mbox{Prob}_\psi\,}
\new{\I}{\mbox{${\cal I}$}}
\new{\lam}{\lambda}
\new{\leta}{\lambda_{\beta}}
\new{\vpsi}{v^\psi }
\new{\vpsit}{v^{\psi _t}}
\new{\lit}{ \lim_{t \to\infty}}
\new{\id}{\mbox{\rm I}}

{\begin{list}{}%
{\settowidth{\labelwidth}{#1}%
\setlength{\leftmargin}{\labelwidth}%
\addtolength{\leftmargin}{\labelsep}%
\setlength{\parsep}{0pt}%
\setlength{\itemsep}{0pt}%
\setlength{\topsep}{0pt}%
}}%
{\end{list}}%
\begin{document}

\thispagestyle{empty}

\begin{center}

{\Large \bf Undergraduate Research Project}\\[5mm]
{ \bf (Proyecto de investigaci\'on)}\\[5mm]

{\bf ​
   Department of Physics\\
   ​Faculty of Exact and Natural Sciences\\
   Universidad del Valle\\
}

\normalsize

\vskip20mm

%%%%%%%%%%%%%%%%%%%%%%%%%%%%%%%%%%%%%%%%%%%%%%%%%%%%%%%%%%%%%%%%%%%%%%%%%%%%%%%%%%%%%%%%%%%%%%%%%%%%%%%%%%%%
%%%%%%%%%%%%%%%%%%%%%%%%%%%%%%%%%%%%%%%%%%%%%%%%%%%%%%%%%%%%%%%%%%%%%%%%%%%%%%%%%%%%%%%%%%%%%%%%%%%%%%%%%%%%
{\Large \bf A BOHMIAN ANALYSIS OF AFSHAR'S EXPERIMENT }
%%%%%%%%%%%%%%%%%%%%%%%%%%%%%%%%%%%%%%%%%%%%%%%%%%%%%%%%%%%%%%%%%%%%%%%%%%%%%%%%%%%%%%%%%%%%%%%%%%%%%%%%%%%%
%%%%%%%%%%%%%%%%%%%%%%%%%%%%%%%%%%%%%%%%%%%%%%%%%%%%%%%%%%%%%%%%%%%%%%%%%%%%%%%%%%%%%%%%%%%%%%%%%%%%%%%%%%%%
% \vfill
\vskip20mm

{\LARGE David Navia\\[10mm]}
% \vfill
\vskip20mm

{\large \rm Supervisor:~Prof.~Ernesto Combariza Cruz~(Departamento de F\'isica, Universidad del Valle)\\
Co-Supervisor:~Prof.~Detlef D\"urr (Mathematisches Institut, LMU, Munich, Germany)}\\[20mm]

{\bf ​
   Santiago de Cali, Colombia\\
   March 18, 2016
   }

\end{center}
\newpage

\tableofcontents

\section*{Abstract}

This work is about \BM, a non-relativistic \qt~about the motion of particles and
their trajectories, named after its inventor David Bohm (Bohm,1952). This
mechanics resolves all paradoxes associated with the measurement problem in
nonrelativistic quantum mechanics. It accounts for quantum randomness, absolute
uncertainty, the meaning of the wave function of a system, collapse of the wave
function, and familiar (macroscopic) reality. We review the purpose for which
Bohmian trajectories were invented: to serve as the foundation of quantum
mechanics, i.e., to explain \qm~in terms of a theory that is free of paradoxes
and allows an understanding that is as clear as that of classical mechanics. To
achieve this we analyse an optical interferometry experiment devised and carried
out 2005 by Shahriar Afshar (Afshar,2005). The radical claim of Afshar implies
in his own words the `observation of physical reality in the classical sense'
for both `which path (particle-like)' and `interference (wave-like)' properties
of photons in the same experimental setup through the violation of the
Englert-Greenberger duality relation (Englert,1996) that according to Englert
can be regarded as quantifying of the \PC.

\section{Introduction}

\bqu{
     ``What exactly qualifies some physical systems to play the role of
     `measurer'?  Was the wavefunction of the world waiting to jump for
     thousands of millions of years until a single-celled living creature
     appeared? Or did it have to wait a little longer, for some better qualified
     system... with a PhD? If the theory is to apply to anything but highly
     idealised laboratory operations, are we not obliged to admit that more or
     less `measurement-like' processes are going on more or less all the time,
     more or less everywhere? Do we not have jumping then all the
     time?'' \cite{Bel90} Bell
}\equ

\bigskip

The classical ideal of passive measurements which simply reveal a preexisting
reality cannot be sustained when we come to a quantum treatment of the
measurement problem. In crude terms, one may say that at the quantum level the
probe becomes as significant as the probed so one cannot `calculate away' its
influence to leave pure information regarding preexisting properties of an
object.  Because these interactions entail transformation of the object that
only in special cases reveal the values of properties without altering them, has
been suggested by Bell the continued use of the word `measurement' in this
context is liable to fuel misconceptions.

In short, Bell argued that the separation between the quantum system and the
measuring apparatus is arbitrary. The encapsulation of the rest of
the world (except the quantum system) into a mathematical entity called an
operator $\hat{G}$, is a very clever trick that allows for straightforward
calculations of the results of quantum measurements without considering
the rest of the world.

In addition to the word `measurement' we have the list of words: `system',
`apparatus', `environment', all this immediately imply an artificial division of
the world, and an intention to neglect, or take only schematic account of, the
interaction across the split. The notions of `microscopic' and `macroscopic'
defy precise definition. So also do the notions of `reversible' and
`irreversible'. Einstein said that it is theory which decides what is
`observable'. We think he was right, and if someone says `observation' our
answer is: Observation of what?.

If one accepts this arbitrary division of the world and that the usual quantum
mechanical description of the state of a quantum system is indeed the complete
description of that system, it seems hard to avoid the conclusion that quantum
measurements typically fail to have results. Pointers on measurement devices
typically fail to point, computer printouts typically fail to have anything
definite written on them, and so on. More generally, macroscopic states of
affairs tend to be grotesquely indefinite, with cats seemingly both dead and
alive at the same time, and the like. This is not good!.
These difficulties can be largely avoided by invoking the measurement axioms of
quantum theory, in particular the collapse postulate, but doing so comes at a
price. One then has to accept that quantum theory involves special rules for
what happens during a measurement, rules that are in addition to, and not
derivable from, the quantum rules governing all other situations.

We believe, however, that the measurement problem, as important as it is, is
nonetheless but a manifestation of a more basic difficulty with standard quantum
mechanics: it is not at all clear what quantum theory is about. Indeed, it is
not at all clear what quantum theory actually says and {\it with what does the
quantum mechanics actually deal ?}, Is quantum mechanics fundamentally about
measurement and observation? Is it about the behavior of macroscopic variables?
Or is it about our mental states? Is it about the behavior of wave functions? Or
is it about the behavior of suitable fundamental microscopic entities,
elementary particles and/or fields? Quantum mechanics provides us with formulas
for lots of probabilities. What are these the probabilities of? Of results of
measurements? Or are they the probabilities for certain unknown details about
the state of a system, details that exist and are meaningful prior to
measurement?

It is often said that such questions are the concern of the foundations of
quantum mechanics, or of the interpretation of quantum mechanics but not. The
problem is not about a philosophical discussion. Philosophy needs physics but
physics needs no philosophy. Physics must be about the objective reality and
need recognizes instead that a quantum theory must describe such a reality. We
need an objective quantum description of nature. We need quantum physics without
quantum philosophy.

This work revolves around an optical interferometry experiment that is a
variation of the Young's experiment. The experiment was devised and carried out
in 2005 by Shahriar Afshar.
Much has been written about this experiment, and from the point of view of the
\oi~of \qm~a satisfactory interpretation of the results could not be found.
Supporters and critics of the Afshar's interpretation of his results have
concentrated on justify or refute the violation of the Englert-Greenberger
duality relation reaching sharply divergent conclusions mutually  incompatible
and contradictory.

This work is only another example that confirms the fact that quantum mechanics,
as Bell quite rightly said, is ``unprofessionally vague and ambiguous.''.
What is usually regarded as a fundamental problem in the foundations of quantum
mechanics, a problem often described as that of interpreting quantum mechanics,
is, we believe, better described as the problem of finding a sufficiently precise
formulation of quantum mechanics, of finding a version of quantum mechanics
that, while expressed in precise mathematical terms, is also clear as physics.
In this work we show with the particular analisys of the \Ae\ `paradox' that
\BM\ provides such a precise formulation.
\section{\Ae}

The \Ae~\cite{Afs05}~uses a setup similar to that for the double-slit
experiment. The aim of the experiment is give information about `which-path' a
particle takes through the apparatus and the contrast or visibility of the
interference pattern (IP) as the quantification of the `wave nature of the
particle'. Afshar claimed that the experiment violates the Englert-Greenberger
relation:

\be
\label{EngIneq}
K^2 + V^2 \le 1
\ee

\begin{figure}[h]
\centering
\includegraphics[scale=0.32]{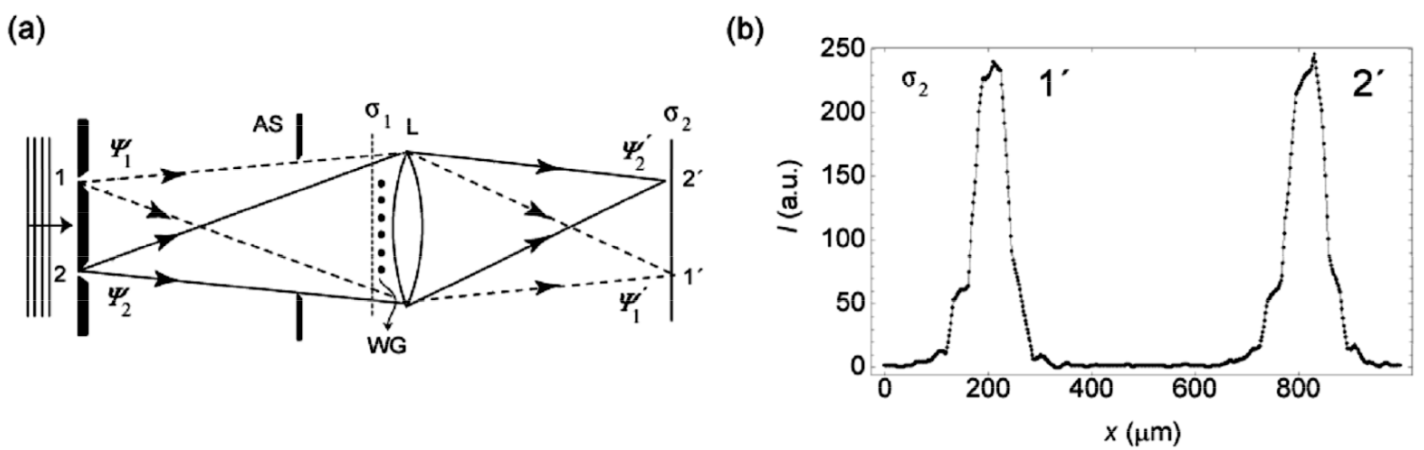}
\caption{(a) Schematics of the Afshar's setup. Laser light impinges upon a dual
        pinhole and two diffracted beams $\psi_1$ and $\psi_2$ emerge. The beams
        are apodized by an aperture stop (AS) and pass through the wire grid
        (WG) and the converging lens ($L$) to produce the irradiance profile in
        the photosensitive surface ($\sigma_2$). (b) Irradiance profile of the
        images at $\sigma_2$, while the WG is present. The irradiance ($I$) is
        measured in arbitrary units (a.u.) of grey-level
        intensity \cite{Afs05}.
        }
\label{fig:Afsharsetup}
\end{figure}

That according to Englert \cite{Eng96} can be regarded as quantifying the notion
of `wave-particle duality' refers to the \PC~\cite{Bohr28}. The stages of the
experiment are:

In stage (i) of the experiment the wire grid (WG) is not present. In this stage
only one slit is left open, therefore, only the corresponding detector gets
illuminated. Opening of the other slit (and illuminating the other detector
simultaneously) does not change the image at the first detector.

In stage (ii) the setup is modified by inserting the WG directly in front of the
lens ($L$).  The WG is carefully positioned in such a way that the wires sit at
the minima of the IP (which can be checked by inserting a screen at the position
of WG and opening the second slit). It is then observed that if one of the two
slits is blocked the image of the remaining open slit is strongly modified due
to the presence of the grid. The wires of the grid reflect (and diffract) some
of the photons, which leads to a reduction of the intensity and introduces the
formation of stripes in the image of the slit.

In stage (iii) the other slit (previously blocked) is reopened (see
Fig.\ref{fig:Afsharsetup}). The crucial result at this stage is that only a
slight reduction in peak intensity is observed at detectors $C_1$ and $C_2$ to
those obtained in the absence of the WG in stage (i).

The experiment has been repeated and its results have been confirmed, concluding
that the fraction of photons that are stopped by the WG is maximally
$1\%$ \cite{AFMK07}. But its interpretation is controversial, and some disagree
that it violates \eq{EngIneq}, while also disagreeing amongst themselves as to
why.

\section{The orthodox origin of the Afhsar's paradox}

\bqu{
    ``Bohr’s principle of complementarity states that quantum systems
    (``quantons'' for short) possess properties that are equally real but
    mutually exclusive. The best known example is what is colloquially termed
    wave-particle duality. In a loose manner of speaking it is sometimes phrased
    similarly to the following: Depending on the experimental situation a
    quanton behaves either like a particle or like a wave'' \cite{Eng96} Englert
}\equ

The radical claim of Afshar \cite{Afs05} implies in his own words the
`observation of physical reality in the classical sense' for both `which-path
(particle-like)' and `interference (wave-like)' properties of photons in the
same experimental setup. His claim of a violation of the inequality \eq{EngIneq}
that according to Englert can be regarded as quantifying the notion of
`complementarity' in wave-particle duality has produced strong criticism
\cite{Kas05,Dre05,Kas09,Ste07,Jac08,Rei08,Qur10,Dre11,Qur12}. Researchers
seeking a `resolution of the paradoxical findings' have resorted to discredit
the validity of Afshar's analysis while also disagreeing amongst themselves as
to why. Below is a brief summary of the papers about the \Ae:

Some authors have tried to argue along the following line. Blocking dark fringes
also blocks out parts of the bright fringes \cite{Jac08,Ste07}. Is that
positioning the WG at the dark fringe locations is not enough to reveal the
existence of the IP, without inducing any further perturbation on the
transmitted light field. This because the WG has an unavoidable effect due to
diffraction, which redirects some light from one `PATH' to the wrong detector.
The introduction of the grid has then according to they partially erased the
`which-path' information since it becomes impossible to univocally associate
each output detector to a given path of the interferometer. They believe, when
both slits are open, the photons contain full `which-path', which is partially
destroyed by partially blocking the bright fringes. They argue that since
without the wires completely blocking the dark fringe, one cannot infer the
existence of the interference, when one tries to increase the information about
the existence of the interference, the `which-path' is proportionately
decreased. In \cite{Flo08} These kind of works have been criticized since assume
the presence of an ideal IP at the WG and then shows using different
experimental situations that the measure of the contrast of the IP is not
possible.

Have been noticed in \cite{Qur12} that in these works, the calculation use as
argument the theoretical existence of full interference, without any blocking
wires. But without the blocking wires, the photons are claimed to have full
`which-path' (`distinguishability = 1' in their language). So, if one were to
take their calculation of `distinguishablity' as correct, then the existence of
full interference in the calculation (in the sense of $|\psi|^2$ yielding an
IP) seems to imply that quantum formalism allows existence of
`interference' for photons for which full `which-path' exists.

Other authors \cite{Kas05,Kas09,Rei08} argues that the `COLLAPSE of the wave
function' TAKES PLACE after the fully articulated `interference' (the
IP is lost DOWNSTREAM from the `which-path' measurement) has
been indirectly indicated to exist by the fact that the grid does not
significantly diminish the intensity of the final detection. But for these
authors the latter measurement doesn’t give any physically meaningful
`which-path' information since the particle already `went through both slits'
and it would not be correct to say that the `which-path' measurement indicates
that the particle `actually went through' only one or the other slit. The use of
different kinds of setup in these works has been criticized in \cite{FK07}.

\iffalse
This is not clear, according to the `COLLAPSE of the wave function', the usual
quantum mechanical dynamics of the state vector of a system given by \se, the
fundamental dynamical equation of quantum theory is abrogated whenever
measurements are performed. The deterministic \se~of the state vector is then
replaced by a random collapse to a state vector that can be regarded as
corresponding to a definite macroscopic state of affairs: to a pointer pointing
in a definite direction, to a cat that is definitely dead or definitely alive,
etc. But this postulate of the random collapse remained silent about how or
where TAKES PLACE.
\fi

Finally in \cite{Dre05,Dre11} is said that the real problem in Afshar's
interpretation comes from the fact that the IP is not actually
completely recorded because to prove experimentally that such sinusoidal
IP actually exists is necessary record photons in the rest of
the wires plane. In \cite{Qur10,Qur12}, is argues that the calculation of
`distinguishability' in \cite{Jac08,Ste07} is fundamentally flawed. This because
if the `state of a particle' is such that the modulus square of the wave
function yields an IP ( and considering that in this
experimental situation the existence of IP is established
without actually disturbing the photons in any way, unlike \cite{Dre05,Dre11}),
then it the detectors detecting the photons behind the converging lens, do not
yield any `which-path' information. This because according to this author `MERE
EXISTENCE of `which-path' information in the state' is sufficient to `DESTROY any
POTENTIAL interference'.

The discussion in this section serves to understand only one thing, the fact
that quantum mechanics, as Bell quite rightly said, is ``unprofessionally vague
and ambiguous'', and that only can produce an immense sea of confusion. This sea
of confusion in this case is composed by the dozens of papers that try to
prevent the violation of the mystic concept of `complementarity',
`complementarity' is only food for mysticism and  only leads to the belief that
the laws of quantum mechanics are a Delphic oracle, which apparently require
high priests to be deciphered, try of understand what `complementarity' is or
even worse defend it, is insane. Only one thing is and can be of course clear
about this discussion and is that defend `complementarity' is defend that the
basic properties of matter can never be understood rationally in terms of unique
and unambiguous models, implies that the use of complementary pairs of
imprecisely defined concepts will be necessary for the detailed treatment of
every domain that will ever be investigated. Thus, the limitations on our
concepts implicit in the \PC~are regarded as absolute and final.

When we said that rational description of nature is possible, we mean that
decoherence does not create the facts of our world, but rather produces a
sequence of redundancies, which physically increase or stabilize decoherence. So
the cat decoheres the atom, the observer decoheres the cat that decoheres the
atom, the environment of the observer decoheres the observer that decoheres the
cat that decoheres the atom and so on.  In short, what needs to be described by
the physical theory is the behavior of real objects, located in physical space,
which account for the facts.

Try to refute Afshar's claim to have shown a violation of `complementarity' by
arguing (using the language of all the previous works) that the `which-path'
information is partially erased or that the contrast of the IP is quite low,
tacitly accepts the notion that \eq{EngIneq} is applicable and assume that
\eq{EngIneq} have some relation with the concept of `complementarity'. Since
Englert makes no mistakes in computation, the first and consequent question is,
whether \eq{EngIneq} applies to the \Ae. In the next chapters is presented a
brief introduction to \BM~ and then is showed from the Bohmian explanation of
the measurement process how this answer can be trivially answered without any
reference to `complementarity'.

\section{\BM}

\BM\, or more exactly `de Broglie-Bohm theory', is a realistic, non-relativistic
quantum theory about moving particles. The Bohmian and the orthodox descriptions
of a measurement produce the same probabilistic predictions. However, the
mathematical implementation of the equations of motion in each case is quite
different. The Bohmian time evolution of the total wave function and the total
trajectory is all we need to explain any quantum process, including the orthodox
measurement process. The orthodox quantum theory requires an operator to
describe the effect of the measuring apparatus, but this operator is not needed
in the Bohmian description of a quantum system of $N$ particles. Here, its {\it
state} is given by variables \cite{DS09}:

\[
	(Q \,,\,\psi)\ ,
\]
where $Q=({\bf Q}_1   \dots    {\bf Q}_N) \in \R ^{3N},$ with
${\bf Q}_k$ being the  positions  of particles, and
$\psi=\psi(q)=\psi ({\bf q}_1   \dots    {\bf q}_N) $ denoting the \wf.\\

We already have an evolution equation for $\psi$, i.e., \Sc's
equation~\cite{DS09}:

\be
\label{eqsc}
i\hbar\frac{\partial \psi }{\partial t}  =
-{\sum}_{k=1}^{N}
\frac{{\hbar}^{2}}{2m_{k}}\Delta_{k}\psi  + V\psi\ ,
\ee the evolution
equation for $Q$ is of the form~\cite{DS09}:

\be
\label{eqvel}\frac {dQ}{dt}=\vpsi(Q)\ ,
\ee where $\vpsi =
({\bf v}^{\psi}_{1}   \dots  {\bf  v}^{\psi}_{N})$.
 Thus, the role
of  $\psi$  is to `choreograph' a motion of particles  through the
vector field on the configuration space that it defines.
By imposing space-time symmetry---Galilean and time-reversal invariance
(or covariance), and `simplicity', we obtain for a general $N-$particle
system~\cite{DS09}:

\be
\label{velo}
{\bf v}_{k}^{\psi} = \frac{\hbar}{m_{k}}{\rm
Im }\frac{ \bold{\nabla}_{k}\psi}{\psi}\ .
\ee

The \wf~affects the behavior of the configuration, i.e., of the particles. This
is expressed by \eq{velo}, but in \BM~there's no back action, no effect in the
other direction, of the configuration upon the \wf, which evolves autonomously
via \eq{eqsc} in which the actual configuration $Q$ does not appear, also for a
multi-particle system the \wf~ $\psi=\psi(q)=\psi ({\bf q}_1 \dots    {\bf q}_N)
$ is not a weird field on physical space, its a weird field on configuration
space, the set of all hypothetical configurations of the system. What it
suggests to us is that you should think of the wave function as describing a law
and not as some sort of concrete physical reality, we want to suggest one should
think about is the possibility that it's nomological \cite{DGN13}, nomic-that
it's really more in the nature of a law than a concrete physical reality.

To connect this theory with the particle mechanics we already
know (Newtonian mechanics), we write the \wf\  in the form $\psi = R\exp[iS/
\hbar]$, where $R$ and $S$ are real. Then the \se\ reduces to the following two
equations~\cite{DS09}:

\be
\label{HamiltonQ}
\frac{\partial S}{\partial t} + \frac{{(\bold{\nabla}S)}^2}{2m} + V + Q = 0\ ,
\ee
where $V$ and $Q$ denote classical and \textit{quantum potential} respectively, such that:

\be
\label{Qpotential}
Q = - \frac{\hbar^{2}}{2m} \frac{\Delta R}{R}\ ,
\ee
and

\be
\label{continuity}
\frac{\partial R^2}{\partial t} + \bold{\nabla} \left (R^2 \frac{\bold{\nabla} S}{m}\right ) = 0 \ .
\ee

Equation \eq{HamiltonQ} is immediately recognized as the classical one-particle
Hamilton-Jacobi equation with an additional term \eq{Qpotential} which vanishes
when $\hbar = 0$~\cite{DS09}. Equation \eq{continuity} is taken to be an
expression for the conservation of probability. Thus, we see that a new quantity
$Q$, the quantum potential, appears alongside classical quantities. It is this
feature that allows us to retain the localized particle with well-defined
positions and momenta, while the novel aspects of quantum phenomena can be
accounted for in terms of the quantum potential~\cite{DS09}.

We've arrived at \BM, defined by equations \eq{eqsc} and \eq{velo} for a
nonrelativistic system (universe) of $N$ particles, without spin. This theory, a
refinement of the Broglie's pilot wave model, was found and compellingly
analized by David Bohm in 1952 \cite{Boh52a,Boh95b}. Spin, as well as Fermi and
Bose-Einstein statistics, can easily be dealt with and in fact arise in a
natural manner.

\BM~is a fully deterministic theory of particles in motion, but a motion of a
profoundly nonclassical, non-Newtonian sort.

\subsection{Quantum equilibrium}

The first question on this matter should be: Which systems should be governed by
\BM ?

Consider an arbitrary initial ensemble $\rho$  and let
$$\rho \to \rho_t$$
be the  ensemble evolution  arising from  Bohmian motion. If
 $\rho= \rho^\psi$  is a functional of $\psi$ we may also consider  the
ensemble evolutions arising from \se\
$$\rho^\psi \to \rho^{\psi_t}\,.$$
 $\rho^\psi$ is {\it equivariant  \/}  if these
evolution are compatible  $$\big(\rho^\psi\big)_t =
\rho^{\psi_t}$$
 That $\rho = |\psi|^2$ is
equivariant follows from comparing the quantum f\/lux equation
\be\label{flux}
 \frac{\partial |\psi|^2}{\partial t} + \hbox{\rm div}\, J^\psi =0
\ee
where $J^\psi = ({\bf J}_1^\psi  \dots  {\bf J}_N^\psi )\, $, $\;
{\bf J}_k^\psi= \frac{\hbar}{m_k}{\rm Im}\ (\psi^*
\nabla_k\psi)$,
with  the continuity equation associated with particle motion
$$\frac{\partial \rho}{\partial t} + \hbox{\rm div}\ \big( \rho \vpsi\big) =0
$$
\smallskip
Since $J^\psi=  v^{\psi}\, |\psi|^2 $, the continuity equation \eq{flux} is satisfied for
$\rho=
|\psi|^2$. Thus:

\bqu
{\it If $\rho(q, t_0) = |\psi(q, t_0)|^2 $ at some time $t_0$  then
$\rho(q, t) = |\psi(q, t)|^2$ for all $t$.  }\equ
 Suppose now that a system has \wf\ $\psi$. We shall call the probability
distribution on configuration space given by $\rho=|\psi|^2$ the {\it
quantum equilibrium} distribution. And we shall say that a system is in
quantum
equilibrium when its configuration are randomly distributed according to the
quantum equilibrium distribution.  The empirical implications of \BM\ are
based on the following \cite{DGN13}

\bqu{{\it Quantum equilibrium hypothesis (QEH): When a system has \wf\
$\psi$, the distribution $\rho$ of its
configuration satisfies $\;\rho = |\psi|^2$}}. \equ

We first remark that it is important recognize that a subsystem need not in
general be governed by \BM, since no wave function for the subsystem need exist.
Thus for a Bohmian universe, it is only the universe itself which a priori,
i.e., without further analysis, can be said to be governed by \BM.  Therefore in
a universe governed by \BM~ there is a priori only one \wf, namely that of the
universe. However, in accordance with this, in a Bohmian universe \cite{DGN13}
\textit{there can be a priori only one system in quantum equilibrium, namely,
the universe itself}. However one question arise from this: Of universes, we
have only one--ours--at our disposal. What possible physical significance can be
attached to a quantum equilibrium ensemble of universes ?

There is a rather simple answer. On the universal level, the physical
significance of quantum equilibrium is as a measure of \textit{typicality}
\cite{DGN13}: for the overwhelming majority of choices of initial $Q$. What we
need to know about, if we are to make contact with physics, is \textit{empirical
distributions}: actual relative frequencies within an ensemble of actual events—
arising from repetitions of similar experiments, performed at different places
or times, within a single sample of the universe -the one we are in-. In other
words, what is physically relevant is not sampling across an ensemble of
universes across (initial) $Q$'s but sampling across space and time within a
single universe, corresponding to a fixed (initial) $Q$ (and $\psi$).

\subsection{Conditional and effective wave function}

The first difficulty immediately emerges: In practice $\rho = |\psi|^2$ is
applied to (small) subsystems. But only the universe has been assigned a wave
function (which we shall now denote by $\Psi$)!. What is meant then by the wave
function of a subsystem?.

Given a subsystem we may write $q=(x,y)$ where $x$ and $y$ are generic variables
for the configurations of the subsystem and its environment. Similarly, we have
$Q_t=(X,Y)$ for the actual configurations (at time $t$). What is the simplest
possibility for the wave function of the subsystem, the $x$-system; what is the
simplest function of $x$ which can sensibly be constructed from the actual state
of the universe at time $t$ (which we remind you is given by $Q_t$ and
$\Psi_t=\Psi$)? Clearly the answer is what we call the {\it \cwf} \cite{DS09}:

$$
\psi(x)=\Psi(x,Y).
$$

This means according to the QEH, that for {\it typical} initial configurations
of the universe, the empirical distribution of  an ensemble of $M$ identical
subsystems with \wf\ $\psi$ converges to $\;\rho = |\psi|^2$ for large $M$.  The
statement refers to an equal-time ensemble or to a multi-time ensemble and the
notion of typicality is expressed by the measure $ \P^{\Psi_0}(dQ)$ and more
importantly by the conditional measure $\P^{\Psi_0}(dQ| \M)$, where the set \M\
takes into account any kind of {\it prior} information---always
present---ref\/lecting the macroscopic state at a time prior to all experiments.
Moreover, the above proposition holds under physically minimal conditions,
expressed by certain measurability conditions ref\/lecting the requirement that
{\it facts} about results and initial experimental conditions not be forgotten.

We remark that even when the $x$-system is dynamically decoupled from its
environment, the conditional \wf\ will not in general evolve according to \Sc's
equation. Thus the conditional \wf\ lacks the {\it dynamical} implications from
which the \wf\ of a system derives much of its physical significance. These are,
however, captured by the notion of \ewf~\cite{DS09}:

Suppose that
\be\label{starstar}
\Psi(x,y)=\psi(x)\Phi(y)+\Psi^\perp(x,y) \,,
\ee
where $\Phi$ and $\Psi^\perp$ have macroscopically disjoint $y$-supports.
If
$$
Y\in \mbox{supp}\,{\Phi}
$$

We say that $\psi$ is the {\it \ewf\/} of the $x$-system. Of course, $\psi$ is
also the \cwf---nonvanishing scalar multiples of \wf s are  naturally
identified. (In  fact, in \BM\ the \wf\  is naturally a projective object since
\wf s differing by a multiplicative constant---possibly time-dependent---are
associated with the same vector field, and thus  generate the same dynamics).

In general systems don't possess and \ewf, for example a system will not have an
\ewf\ when, for example, it belongs to a larger microscopic system whose \ewf\
doesn't factorize in the appropriate way. However, the {\it larger} the
environment of the system, the {\it greater} is the potential for the existence
of an \ewf\ for this system, owing in effect to the abundance of
`measurement-like' interactions with a larger environment.

There is a natural tendency toward the formation of  stable  \ewf s  via {\it
dissipation}: Suppose that initially the $y$-supports of $\Phi$ and
$\Psi^{\perp}$ are just `sufficiently' (but not macroscopically) disjoint; then,
due to the interactions with the environment, the amount of $y$-disjointness
will tend to increase dramatically as time goes on, with, as in a chain
reaction, more and more degrees of freedom participating in this disjointness.
When the effect of this dissipation, are taken into account, one find that even
a small amount of $y$-disjointness will often tend to become `sufficient,' and
quickly `more than sufficient,' and finally macroscopic.

The ever-decreasing possibility of interference between macroscopically distinct
\wf s due to typically uncontrollable interactions with the environment it is
what is known in \oqm\ as {\it decoherence}.

\subsection{Quantum randomness and absolute uncertainty}

Absolute uncertainty is a consequence of the analysis of $\;\rho = |\psi|^2$.
It expresses the impossibility of obtaining information about positions more
detailed than what is given by the quantum equilibrium distribution. It provides
a precise, sharp foundation for the uncertainty principle, and is itself an
expression of global quantum equilibrium. This is that the quantum equilibrium
hypothesis $\;\rho = |\psi|^2$ conveys the most detailed knowledge possible
concerning the present configuration of a subsystem \cite{DGN13}.

In general quantum randomness and absolute uncertainty are merely an expression
of quantum equilibrium, a global configurational equilibrium subordinate to the
universal (and, in fact, nonequilibrium) wave function $\Psi$. More exactly
absolute uncertainty in \BM~justifies the dual role of $\psi$: it has, in
addition to its statistical aspect, also a dynamical one, as expressed in
\eq{eqsc} and \eq{eqvel}. Thus, knowledge of the wave function of a system,
which sharply constrains our knowledge of its configuration, is knowledge of
something in its own right, something real and not merely knowledge that the
configuration has distribution $|\psi|^2$.
\subsection{Particle trajectories in the double-slit experiment}
\label{subsec:doubleslit}

\begin{figure}[h]
    \centering
    \includegraphics[scale=0.38]{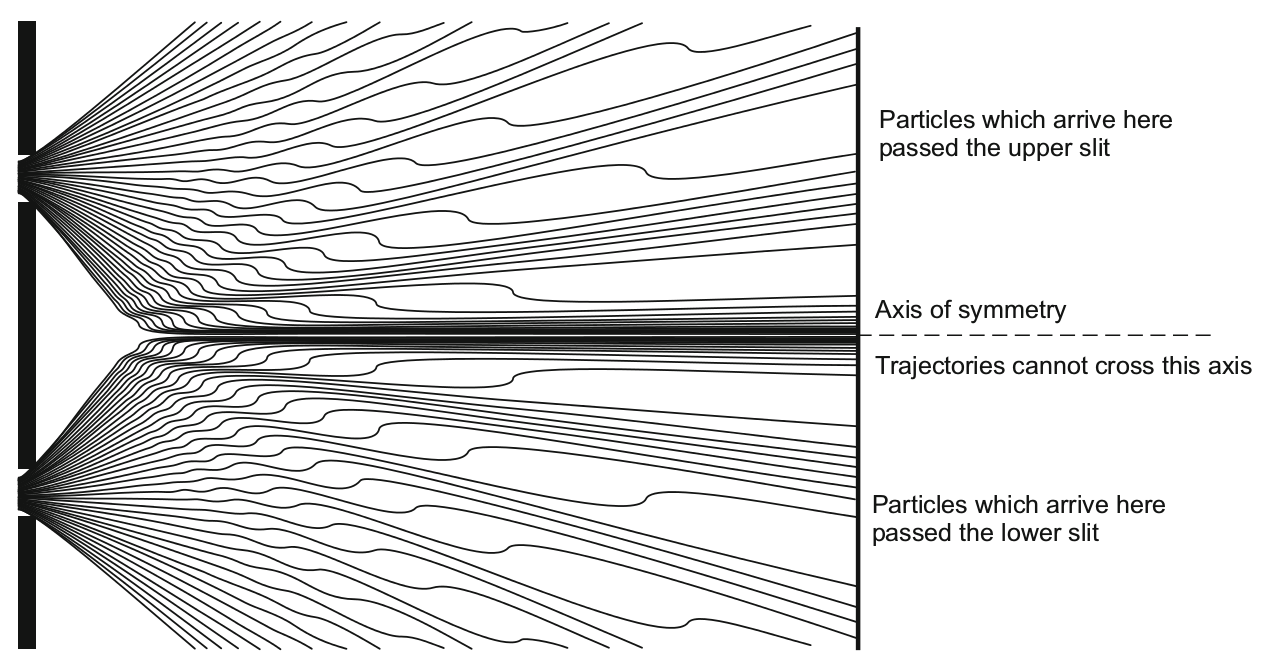}
    \caption{
    Possible trajectories for a particle in the famous double slit
    experiment \cite{Phi79}.
    After passage through the slits, the guiding wave $\psi$ of every particle
    forms a diffraction pattern according to Huygens' principle. After the slit,
    this diffracted wave determines the possible trajectories of the
    particle.
    }
    \label{fig:trajec}
\end{figure}

In \BM~the wave always goes through both slits and the particle goes through
only one. But the particle is guided by the wave toward places where $|\psi|^2$
is large, and away from places where is small. As any wave would do, after
passage through the slits, the guiding wave $\psi$ of every particle forms a
diffraction pattern and so if a plate is in position the particle contributes a
spot to the IP which is essentially the quantum equilibrium distribution on the
plate at their point of arrival, this is clear enough, because that is what the
quantum equilibrium hypothesis says. In no case the earlier motion, of either
particle or wave is affected by the later insertion or non insertion of the
plate (see Fig.\ref{fig:trajec}).

This result is clear, but the double slit experiment is often taken in textbooks
as proof that one cannot have moving particles in quantum physics.

In general is not useful compute Bohmian trajectories in various quantum
mechanical situations, what we learn in general from the trajectories is that,
at each moment of time $t$, the particles have positions which are distributed
according to the quantum equilibrium hypothesis. In \BM\ we can make more
detailed predictions about the behaviour of individual elements in the ensemble.
We cite 4 examples of predictions pertaining to individual particles in the
double slit experiment \cite{DS09}:

\begin{enumerate}
    \item The trajectories cannot cross the axis of symmetry.
    \item \label{itm:predic2}The trajectories move mostly along the maxima (hyperbola) of $|\psi|^2$
    and spend only short times in the valleys of $|\psi|^2 (|\psi|^2 \approx 0)$.
\item The trajectories cross the valleys since, right after the slits, the
    trajectories expand radially. This is turn happens because the guiding wave
    is a spherical wave close behind the slits, and while first feeding the
    nearest maxima, they must observe quantum equilibrium. Most trajectories
    will have to lie in the region of the main maximum (around the symmetry
    axis, which is the most clearly visible on the screen), i.e., trajectories
    must cross over from adjacent maxima to the main maximum.
\item  The arrival spot on the screen is random, in particular the slit through
    which each particle goes is random. The randomness is due to the random
    initial position $Q_0$ of the particle with respect to the initial wave
    packet.  By always preparing the same wave packet $\psi$, one prepares an
    ensemble of $|\psi|^2$ distributed positions.

\end{enumerate}

We emphasize that these predictions concerning individual behaviour do not
contradict the statistical ones of \qm. Their purpose is simply to contribute
to the explanation of how the latter come about.

\subsection{Delayed-choice experiments from the Bohmian perspective}
\label{subsec:delayedchoice}

\bqu{
    ``Is it not clear from the smallness of the scintillation on the screen that
    we have to do with a particle? And is it not clear, from the diffraction and
    interference patterns, that the motion of the particle is directed by a
    wave? De Broglie showed in detail how the motion of a particle, passing
    through just one of two holes in screen, could be influenced by waves
    propagating through both holes. And so influenced that the particle does not
    go where the waves cancel out, but is attracted to where they cooperate.
    This idea seems to me so natural and simple, to resolve the wave-particle
    dilemma in such a clear and ordinary way, that it is a great mystery to me
    that it was so generally ignored.''  \cite{Bel87} Bell
}\equ

The traditional demonstration that one cannot simultaneously observe the `path'
a particle takes through an interferometer and the IP it contributes to rests on
an application of the Heisenberg relations. It has not been generally
appreciated that the desired result may be arrived at without any direct
invocation of Heisenberg's relations, but follows from general properties of the
many-body interaction between a coherent system and a detector. This seems to
provide a deeper understanding of the problem in that applications of
Heisenberg's relations have tented to rely on arguments drawn from classical
optics whose connection with the quantum formalism is not always clear. We shall
see that even if the coherence of the interfering beams is not destroyed by the
external path-measuring agency, interference will still not be observed in
situations where we determine the `path'. Actually, the usual approach to this
problem, however it proceeds, is deficient in one fundamental respect

\bqu{
    It seeks to define the conditions under which one can or cannot have
    definite knowledge about the `path' a quantum particle takes while working
    with the formalism of \oqm~within which such a notion is meaningless in all
    circumstances.
}\equ

To discuss this problem consistently means admitting that $\psi$ provides an
incomplete description of individual events and must be supplemented by the idea
that in our world, electrons and other elementary particles have precise
positions ${\bf Q}_k(t) \in \R^3$ at every time $t$ that move according to
\eq{velo}. That is, for a certain trajectory $t \mapsto Q(t)$ in configuration
space, it claims that $Q(t) = ({\bf Q_1}(t),\cdots,{\bf Q_N}(t))$ is the
configuration of particle positions in our world at time $t$.

This, of course, is precisely what \BM~provides and it is our purpose here to
put on a secure conceptual footing the proof of the impossibility of
simultaneous measurement of a particle position and an interference image
(`simultaneous' means that in an individual case we obtain a particle position
and contributes to an IP generated by the overlap of two coherent waves which is
essentially the quantum equilibrium $|\psi|^2$ distribution). Notice that we
don't say measurement of the trajectory since as consequence of the uncertainty
principle is impossible measure the trajectory of an individual particle takes,
because any measurement of position irrevocably disturbs the momentum, and vice
versa.  Using weak measurements, however, it is possible to operationally define
a set of trajectories for an ensemble of quantum particles \cite{KBRSMSS11}.

We also show that the trajectory assumption is not disproved by the traditional
argument and on the contrary, it serves to justify the latter. Remember that we
consider the \wf~as nomological and is still relevant when we possess the
particle position because particles move in a way that depends on the \wf.
Notice that the notion of `wave-particle duality' refers to `complementarity'
does not apply in our model.

Consider a beam of particles (normalized wave packet $\psi$) sent through on a
beam splitter which produces two ensembles of independent subsystems (two
packets $\psi_1$ and $\psi_2$) that propagate along the arms of an
interferometer. A device designed to measure the position of the particle
($x$-system) is situated in the upper arm. The meter needle of this device has
initial coordinate $y_0$ in a normalized packet $\phi_{y_0}$. It is assumed that
$\psi_1 \cap \psi_2 = 0$ in order that the observing apparatus interacts only
with $\psi_1$.

According to orthodox quantum measurement theory, after a measurement, or
preparation, has been performed on a quantum system, in this case the
$x$-system, the wave function for the composite formed by system and apparatus
is of the form \cite{DGN13}

\be
\label{omeasur}
\sum \limits_{\alpha} \psi_{\alpha} \otimes \phi_{\alpha}
\ee

With the different $\phi_{\alpha}$ supported by the macroscopically distinct
(sets of) configurations corresponding to the various possible outcomes of the
measurement, e.g., given by apparatus pointer orientations. Of course, for
\BM~the terms of \eq{omeasur} are not all on the same footing: one of them, and
only one, is selected, or more precisely supported, by the outcome,
corresponding, say, to $\alpha_0$ which actually occurs. To emphasize this we
may write \eq{omeasur} according to \eq{starstar} in the form

\[
\psi \otimes \phi + \Psi^{\perp}
\]

Where $\psi = \psi_{\alpha_{0}}$, $\phi = \phi_{\alpha_0}$, and $\Psi^{\perp} =
\sum \limits_{\alpha \not= \alpha_0} \psi_{\alpha} \otimes \phi_{\alpha}$. It
follows that after the measurement the $x$-system has effective wave function
$\psi_{\alpha_0}$. We can determine the position of the particle by performing a
sharp position measurement. However, a position determination only requires a
`weaker measurement' which locates the particle in the region of space where
$\psi_1$ is finite. Hence, we assume that the interaction with the observing
device does not appreciably alter $\psi_1$.

The aim in this case is to maximize the perturbation of the $x$-system and
minimize that between the initial and final states of the $y$-system
(apparatus). We shall show that these are mutually incompatible requirements.
At the exit to the beam splitter the total wave function is

\[
\psi_0 = [\psi_1 + \psi_2]\phi_{y_0}
\]
\[
\rightarrow f + \psi_2\phi_{y_0}\
\]
\[
\rightarrow \psi = \psi_1\phi_{y} + \psi_2\phi_{y_0}
\]

After the interaction, where $\phi_{y}$ is the final state of the meter and
where, as we have said, $\psi_1$ is essentially unaltered. If $\psi_1$ and
$\psi_2$ still do not overlap, the configuration space summands in will not
overlap and the system point to which $x$-system belongs is in one of them.
However, in order to say unambiguously which one from observation of the various
possible outcomes of the measurement, we must require that $\phi_{y_0}$ and
$\phi_y$ are disjoint $(\phi_{y_0} \cap \phi_{y} = 0)$, that is the initial and
final apparatus states must be orthogonal. The position of the particle is then
determined by the distinguishable outcomes of the meter and we may infer
the `path' from how the coordinate $y$ changes

\[
    \begin{cases}
        x \in \psi_1\ \mbox{iff} \ y_0 \rightarrow y \in \phi_y \\
        x \in \psi_2\ \mbox{iff} \ y_0 \rightarrow y \in \phi_{y_0}
    \end{cases}
\]

But, under these circumstances, no interference will be observed when $\psi_1$
and $\psi_2$ subsequently overlap, whatever route the particle took

\be
\label{Intensity}
I = {|\psi_1|}^2 + {|\psi_2|}^2 + [c\psi_2^{*}\psi_1 + cc]
\ee

Where

\be
\label{InterConst}
c = \phi^{*}_{y_0}\phi_{y}
\ee

Clearly, if $c=0$ the interference terms vanish. We conclude that a
determination of the position of a particle in an interferometer is incompatible
with the observation of interference. The general principle at work that causes
the pattern to be washed out is clear, the mere fact that there is an exchange
of energy or momentum between the system and apparatus. The nub of the issue is
that in order to perform its role and unambiguously reveal a change, the initial
an final states of the observing device must be orthogonal. Between the extremes
of $\phi_y = \phi_{y_0}$ (maximum contrast) and $\phi_{y_0} \cap \phi_y = 0$ (no
interference) there is a continuous range of diminutions in the contrast of the
IP depending on the extend of the overlap of $\phi_{y_0}$ and $\phi_y$.  Then is
impossible obtain an exact position of the particle, except when $\phi_{y_0}
\cap \phi_y = 0$. Our argument against the possibility of a simultaneous
observation of position and interference is evidently of a general character and
not dependent on the nature of the interacting systems.  Notice also that we
arrive at this conclusion without invoking the `collapse of the wave function'
on `measurement'.

\begin{figure}[h]
    \centering
    \includegraphics[scale=0.35]{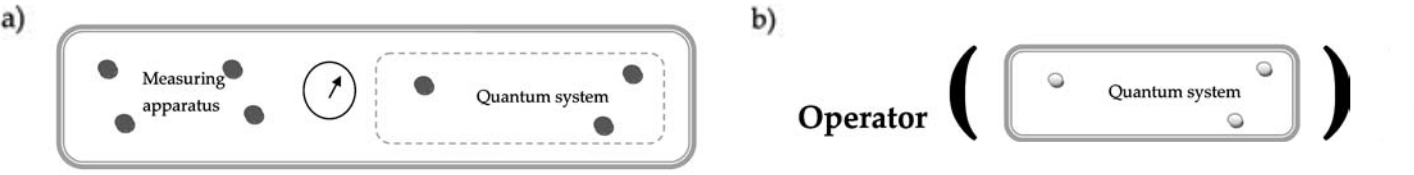}
    \caption{
    (a) A Bohmian measurement assumes that the quantum system and the measuring
    apparatus are explicitly simulated \cite{OM12}. (b) The orthodox
    measurement assumes that only the quantum system is explicitly simulated,
    but the measuring apparatus is substituted by an operator acting on the wave
    function of the system.
    \label{fig:measurement}
    }
\end{figure}

This is how collapse (or reduction) of the effective wave function to the one
associated with the outcome $\alpha_0$ arises in \BM. While in \oqm~the
`collapse' is merely superimposed upon the unitary evolution without a precise
specification of the circumstances under which it may legitimately be invoked,
we have now, in \BM, that the evolution of the \ewf~is actually given by a
stochastic process, which consistently embodies both unitarity and collapse
clearly and unambiguously (see Fig.\ref{fig:measurement}).
In particular, the \ewf~of a subsystem evolves according to \se~when this system
is suitably isolated. Otherwise it ``pops in and out'' of existence in a random
fashion, in a way determined by the continuous (but still random) evolution of
the \cwf~$\psi_t$ \cite{DGN13}. Moreover, it is the critical dependence on the
state of the environment and the initial conditions which is responsible for the
random behavior of the (conditional or effective) wave function of the system.

Because in \BM~the \wf~is more than a mere repository of our current information
about a system, and its law of motion induces that of the particle, it is a
natural to expect that the insertion of a probe in the apparatus to check the
above predictions will modify the wave in such a way that the particle
trajectory suffers a disturbance.  It has been proposed that such experiments
allow one to glean varying degrees of `knowledge' of the `particle and wave
aspects of matter'. Is necessary clarify the significance has been attributed in
the literature to the circumstances that the fringes may remain substantially
visible when the `particle knowledge' is relatively large. This is a simple
consequence of the definition of the contrast of an interference pattern
\cite{Hol93}.

Let $\psi = \psi_B + \psi_{B'}$, where $\psi_b,\psi_{B'}$ are arbitrary waves of
amplitude $R_b, R_{B'}$, and write $I = |\psi|^2$. Then the contrast, which has
been presented as a measure of our `wave knowledge', is defined by:

\be
\label{contrast}
V = \frac{I_{max} - I_{min}}{I_{max} + I_{min}} = \frac{2R_BR_{B'}}{R_B^2 + R_{B'}^2}
\ee

With $0 \leq V \leq 1$. As the ratio $\frac{R_{B'}}{R_{B}}$ decreases, $V$
decreases but at a slower rate. To bring out the implications of this define a
`complementary' quantity characterizing the `particle knowledge'

\be
\label{complement}
K = \frac{R_B^2 - R_{B'}^2}{R_{B}^2 + R_{B'}^2}
\ee

We have

\be
\label{dualrel}
K^2 + V^2 = 1
\ee

The problem with this discussion is that for it to be meaningful we must give
credence to the notion that a particle has a `path' through the interferometer
that we could have `knowledge' of. This concept has been slipped in without an
explicit recognition that it is an additional assumption. In this fact, only in
\BM~that the `which-path' debate fully makes sense.

The mathematical discussion presented above does not require \qm~at its heart.
In particular, the derivation of \eq{dualrel} is essentially in terms of the
diffraction and interference of waves valid for waves of any sort. In this
connection one must remember that $V$ and $K$ defined above are primarily
expressions reflecting properties of the interfering waves and their variation
represents objective changes in the system (note that in general they are
functions in space). As secondary properties is possible apply this result to
the \wf~of a quantum system to yield statistical information of it. In that case
\eq{dualrel} is known as the Englert-Greenberger duality relation \eq{EngIneq}.

\section{\Ae~again}

The discussion around if the \Ae~violates \eq{EngIneq} and so the \PC~is kind of
silly. The \PC~is ill defined and this makes the critical analysis difficult.
What we mean is this: wave and particle are words, they have of course through
everyday language a clear connotation but since we are doing physics, these
words have to be linked to physics, i.e., to a physical theory. Now in the \Ci,
these linkage is not done via a theory but through experimental setups. Those
experimental setups are supposed to `define' what particle and what wave mean.
That is of course delicate and obscure, and that is the problem with the \PC.
Suppose you have an experiment which defines `particle', for example because the
experimental data are of position measurement and give the statistics as they
are given by the amplitude square of the wave function. Then suppose you have
another experiment which measures `wave' (no such experiment exists!). Usually
people think of IP in the modulus square of the wave function as such. Then
combining such experiments one may derive some kind of relations like
\eq{EngIneq}, but that relation holds then only for that kind of setup. It is a
theoretical description of some kind of experimental setup. But through those
words like `which-path' and `interference' such relations sound interesting. It
is the way people talk without reference to a physical theory that things look
interesting or mysterious.

Englert uses for the derivation of \eq{EngIneq} in \cite{Eng96} an experimental
setup analogous to the setup that we discussed in the section
\ref{subsec:delayedchoice}. We have shown in that section from the Bohmian
perspective the explanation of why a determination of the position of a particle
in an interferometer is incompatible with the observation of interference, i.e.,
that in an individual case if we obtain the particle position that particle
can't contributes to an IP generated by the overlap of two coherent waves.

The \Ae~setup is a very different kind of setup, in stage (iii) the contrast is
inferred using \eq{contrast}, which is essentially measure the minimum intensity
of the dark fringe $I_{min} \approx 0$ from comparative measurements of the
total flux with and without the WG \cite{AFMK07}:

\[
V = \frac{I_{max} - I_{min}}{I_{max} + I_{min}} = \frac{I_{max}}{I_{max}} = 1
\]

This process clearly doesn't affect the quantum equilibrium distribution and
allows to perform the `which-path' measurement that as we explained in the last
section is meaningless in terms of the \oi~of \qt. With this in mind now
something is clear, The \Ae~correspond to a very different physical process for
which \eq{EngIneq} was not compute and therefore doesn't apply.

The characterization of the `wave and particle aspects of matter' in the
\oi~does not exhaust all we can say about these concepts and does not reflect
the meaning given to them in \BM. In particular the function $K$ in \eq{EngIneq}
and \eq{complement} has no connection with the Bohmian trajectories. The fact is
that \oqm~not simply a theory of our knowledge of physical systems, a kind of
generalization of classical statistical mechanics without the trajectories, and
diffraction phenomena demonstrate this.

We hope that our analysis can contribute to clarify the confused debate
originated by the orthodox analysis of the \Ae. While this results may not
appear as a big surprise the \Ae~is not trivial. We have shown with the previous
discussion an example of the paradoxes that may arise if one attempts to
artificially subdivide quantum phenomena and pretend to study it adopting an
inconsistent model to describes what a quantum particle is and how it behave.
To avoid the paradoxical conclusion of the \oi~of \qt~that apparently flowing
from the `delayed-choice' class of experiments, i.e., that the decision on
whether to insert the counters (and hence determine `which-path') or to leave
them out and let the electron contribute to the IP (the electron takes `both
paths') may be made \textit{after} the electron has `passed' the slit plane and
so the that the earlier behaviour of the electron (passage through one or both
slits) may be influenced by our later decision whether or not to insert the
counters. It is proposed by Wheeler against this nonsense and in defence of a
`consistent' \oi, that matter has disappeared!! and so to be is to be
perceived!!, he himself seems to be very clear on this point.  He says:

\bqu{
    ``After the quantum of energy has already gone through the doubly slit
    screen, a last-instant free choice on our part--we have found--gives at will
    a double-slit-interference record or a one-slit-beam count. Does this result
    mean that present choice influences past dynamics, in contravention of every
    formulation of causality? Or does it mean, calculate pedantically and don't
    ask questions?  Neither; the lesson presents itself rather as this, that the
    past has no existence except as it is recorded in the present. It has no
    sense to speak of what the quantum of electromagnetic energy was doing
    except as it is observed or calculable from what is observed.'' \cite{Whe78}
}\equ

This conclusion is often cited as conflicting with the idea that there can be
particles with trajectories(or more general as conflicting with the idea of the
existence of an external real world out there and that it is the task of physics
to find the basic constituents of this exterior real world and the laws that
govern them). One sends a particle through a double slit (i.e., a wave packet
$\psi$). Behind the slit at some distance is a photographic plate. When the
particle arrives at the plate it leaves a black spot at its place of arrival.
Nothing yet speaks against the idea that the particle moves on a trajectory. But
now repeat the experiment. The next particle marks a different spot of the
photographic plate.  Repeating this a great many times the spots begin to show a
pattern. They trace out the points of constructive interference of the wave
packet $\psi$ which, when passing the two slits, shows the typical Huygens
interference of two spherical waves emerging from each slit. Suppose the wave
packet reaches the photographic plate after a time $T$(this time is defined in
terms of the motion of the centre of a wave packet). Then the spots show the
$|\psi(T)|^2$ distribution, in the sense that this is their empirical
distribution. Analyzing this using \BM~mechanics, i.e., analyzing \se~and the
guiding equation, one immediately understands why the experiment produces the
result it does. It is clear that in each run the particle goes either through
the upper or through the lower slit. The wave function goes through both slits
and forms after the slits a wave function with an IP. Finally the repetition of
the experiment produces an ensemble which checks Born's statistical law for that
\wf. That is the straightforward physical explanation.  So where is the argument
which reveals a conflict with the notion of particle trajectories? Here it is:

From equations \eq{Intensity} and \eq{InterConst} we can observe that in
general:

\be
\label{decoh}
I = {|\psi_1|}^2 + {|\psi_2|}^2 + [c\psi_2^{*}\psi_1 + cc] \not= {|\psi_1|}^2 + {|\psi_2|}^2
\ee

Wheeler arrives at his conclusion invoking \eq{decoh}, one of the apparent
non-localities of \oqm, i.e., the instantaneous, over all space, `collapse of
the wave function' on `measurement' \cite{Bel87}.

The \Ae~although not violate \eq{EngIneq}, show a very important result, Afshar
observes that the amount of light intercepted by the WG is very small, is
maximally $1\%$ \cite{AFMK07}. This is a clear experimental proof of the
prediction \ref{itm:predic2} that \BM~makes concerning to the double-slit
experiment that we reference in the section \ref{subsec:doubleslit} and more
important, this mean that \eq{decoh}, i.e., the mathematical expression of
`decoherence', can't give a complete account of the quantum process and so is
necessary a theory that provide us with a coherent story about the complete
quantum process itself.  This, of course, is precisely what \BM~provides,
interpreting the problem simply as evidence of the incompleteness of \qt. The
passage of the corpuscle through one slit and the wave through both forms a
well-defined time-dependent physical process in itself. What happens
subsequently has no bearing on it at all. If both paths through the
interferometer are open the particle will respond to the overlapping waves. The
detecting plate reveals that this has happened, but does not influence it. If
instead the counters are inserted prior to the overlap of the waves the evolved
wave is different and in detecting the particle the counters again simply reveal
this. According to \BM, \textit{The present merely reveals the past and has no
influence upon it}. In this regard at least there is no need to revise our
customary conceptions of cause and effect, nothing simply surges up out of
nothing without having antecedents that existed before.  Likewise, nothing ever
disappears without a trace, in the sense that it gives rise to absolutely
nothing existing at later times. This principle is not yet a statement of the
existence of causality in nature. Indeed, it is even more fundamental than is
causality, for it is at the foundation of the possibility of our understanding
nature in a rational way \cite{Boh71}.

\section{Bohmian trajectories as the foundation of quantum mechanics}

The core of \oqm~is that physics is about measurement and observation, and not
about an objective reality, about what seems, i.e., `observation' or
`sensations' or `information' or `data' and not about what is about the
things-in-themselves!.

We think that a fundamental physical theory should not be formulated in terms of
concepts like `observer' or `experiment,' as these concepts are very vague and
certainly do not seem fundamental. Is a cat an observer? A computer? Were there
any experiments before life existed on Earth? If consciousness causes collapse
and consciousness is a product of the brain What about the collapse of the
brain? Does man think with the help of the brain?

In this work we have studied only a simple example of the crisis in quantum
physics, but the essence of the crisis consists in the rejection of an objective
reality existing outside the mind. `Matter has disappeared' one may thus
express the fundamental and characteristic difficulty in relation to many of the
particular questions, which has created this crisis. The idea that now dominates
the minds of physicists is that laws of motion can be treated mathematically and
therefore that mathematics can to overlook matter. `Matter disappears,' only
equations remain.

In this new stage of development and apparently in a new manner, we get the old
Kantian idea: reason prescribes laws to nature. This, of course, is a ridiculous
dream. `Matter is disappearing' means that the limit within which we have
hitherto known matter is vanishing and that our knowledge is penetrating deeper;
properties of matter are likewise disappearing which formerly seemed absolute,
immutable, and primary and which are now revealed to be relative and
characteristic only of certain states of matter.

\bqu

The {\it sole property of matter} that necessarily must be recognized by physics
to advance is the property of being an {\it objective reality}, of {\it existing
outside our mind}.

\equ

In a nutshell, the quantum philosophy of today is the same `philosophical
idealism' of yesterday, this means that one school of scientists in one branch
of natural science has slid into an absurd philosophy. The abandonment of these
ideas is being made, and will be made, by quantum physics; but it is making not
directly, but by zigzags, not consciously but instinctively, not clearly
perceiving its `final goal,' but drawing closer to it gropingly, hesitatingly,
and sometimes even with its back turned to it. Quantum physics is in travail; it
is giving birth to future. The process of child-birth is painful. And in
addition to a living healthy being, there are bound to be produced certain dead
products, refuse fit only for the garbage-heap. And the entire school of
positivism and empirio-critical philosophy, together with empirio-symbolism,
empirio-monism, and so on, and so forth, must be regarded as such refuse!

Bohmian trajectories have been used for various purposes, including the
numerical simulation of the time-dependent Schrodinger equation and the
visualization of time-dependent wave functions \cite{WT05}. But we have showed
in a simple example their real purpose they were invented for: to serve as the
foundation of quantum mechanics, i.e., to explain quantum mechanics in terms of
a materialistic theory, providing the foundation for the claim that a
fundamental physics focused on an objective reality existing outside the mind is
possible, providing a theory free of paradoxes and allows an understanding that
is as clear as that of classical mechanics, this means, while expressed in
precise mathematical terms is also clear as physics.

One of the most absurdity objections to \BM~is that is more complicated than
orthodox quantum theory, since it involves an extra equation and is not
`economical'. One has only to put this objection in order to see the absurdity,
the subjectivism of applying the category of `the economy of thought' here.
Human thought is `economical' only when it correctly reflects objective truth,
and the criterion of this correctness is practice, experiment and
industry \cite{BASMO14}, all mysteries which lead a theory to mysticism find
their rational solution in human practice and in the comprehension of this
practice.

\BM~emerges as a precise and coherent quantum theory providing a microscopic
foundation for the quantum formalism. To sum up, it seems fair to say that
\BM~is not an interpretation of anything it is a complete theory where nothing
is left open, and above all, it does not need an interpretation. It is a theory
of nature, and it has a precise link to quantum mechanics. At the end of the
day, the fundamental reasons for the dismissal seems to be this: \BM~is against
the spirit of quantum mechanics. That is, an objective physical description, a
return to determinism, a return to physical clarity clash with the tenets of
`philosophical idealism'.

\section{Towards a nonlinear quantum physics}

\subsection{The classical limit in \BM}

As could be seen \BM~provides a  precise and coherent quantum theory with a
microscopic foundation for the quantum formalism, giving a rigorous mathematical
description together with a clear and unambiguous picture of the motion of
quantum objects, in the non-relativistic regime.

Newtonian mechanics smoothly arises from special relativity as the speed of the
system becomes smaller and smaller than the speed of light. This is possible
because both theories are built upon the same conception of systems as localized
(point-like) objects in space and time.

Similarly in quantum mechanics, according to standard textbook the criteria for the
classical limit, it is often considered  when $\hbar \to 0$ (or equivalently
$\hbar \ll S_{cl}$ , where $S_{cl}$ denotes a related classical action).
Therefore, in Bohmian mechanics everything should be very simple: the condition
$\hbar \to 0$ should lead to classical mechanics. Bohmian mechanics has a very
appealing feature regarding the quantum to classical transition: the quantum
potential. This additional term to the Hamilton-Jacobi equation contains
information about the topological curvature of the wave function in the
configuration space and is proportional to $\hbar^2/m$:

\[
    m\frac{d^2\textbf{X}}{dt^2} = \textbf{F} + \textbf{F}_Q
\]

\[
    m\frac{d^2\textbf{X}}{dt^2} \simeq \textbf{F}
\]

Where $\textbf{F} = -\bold{\nabla}V(\textbf{X})$ and $\textbf{F}_Q =
-\bold{\nabla}V_Q(\textbf{X})$ are respectively the classical force and
the ``quantum'' force of a particle with trajectory $\textbf{X} =
\textbf{X}(t)$.

This is the traditional classical limit. Now, $\hbar$ is just a constant and
therefore is also possible think the classical limit in a more physical way,
in terms of larger and larger masses.

Now, what happpend in this case? The quantum potential should also vanish and
therefore the classical Hamilton-Jacobi equation should start ruling the system
dynamics. What we observe as $m$ increases is that trajectories start mimicking
the behavior of classical trajectories. One could be tempted to think that we
are actually observing the emergence of classical trajectories. However, because
of the intricate interference structure of the wave function, the quantum
potential also exhibits a rather complex topology, such that the factor
depending on the curvature of the wave function does not vanish. Consequently,
one of the distinctive traits of Bohmian mechanics, the noncrossing rule, still
remains valid.

\iffalse
To properly address the question of the classical limit or, first it is worth
noting that \BM~is confined to the configuration space (as it is
the wave function, unless other representation is chosen). However, classical
mechanics takes place in phase space, where coordinates and momenta are
independent variables, and therefore at each time it is possible to observe that
the same spatial point can be crossed by trajectories with equal but opposite
momenta.

This is a strong condition that should be satisfied in a proper classical limit;
reaching the classical Hamilton-Jacobi equation is just a weak condition, as
seen above, since it does not warrant the observation of twofold momenta. So,
what can we do in this situation?  Is there any way to get rid of the fast
oscillations that appear in the so-called classical limit (independently of how
this limit is reached)?
\fi

The key ingredient in the analysis of the emergence of the classical world in
\BM~is then the following: If one observes the dynamics of the full-dimensional
system (system of interest plus environment), the corresponding Bohmian
trajectories satisfy the usual noncrossing rule. Now, these trajectories contain
information about both the system and the environment.

In order to examine the system dynamics, one has to select only the respective
components of those full-dimensional trajectories. This is equivalent to
observing the dynamics in a subspace, namely the system subspace. It is in this
subspace where the system trajectories display crossings, just because they are
not (many-particle) Bohmian trajectories in the high-dimensional space, but
trajectories onto a particular subspace, actually this leads us immediately to
the notion of conditional wave function and his possible use to establish a
bridge between these various subspaces. Here the role of the environment
consists of relaxing the system noncrossing property by allowing its (reduced)
trajectories to reach regions of the configuration subspace which are
unaccessible when the system is isolated. This thus explains the phenomenon of
decoherence.

In order to discuss the practical utility of these conditional wave functions,
we need to explore the possibility of defining them independently of the big
wave function, ie, find an equation to compute and analyse the dynamical
evolution of a quantum (sub)system exclusively in terms of these single-particle
(conditional) wave functions, instead of (as we did in their presentation) first
solving the many-body \se~for $\Psi$ and only examining the conditional wave
functions $\psi$ after.

However, contrarily to the \se~the new equation can be nonlinear. In other
words, there is no guarantee that the superposition principle satisfied by the
big wave function (in the big configuration space) is also applicable to the
little (conditional) wave function when dealing with quantum subsystems.

This analysis is interesting in the context of collapse models. In
\cite{Toros16} it is shown that the dynamics of the \cwf~obeys to a collapse
type of equation which, under reasonable assumptions, is that of the GRW model.
In addition, for the center of mass the \cwf~of a compositive system, an
amplification mechanism is present, like in the GRW model.

\subsection{Collapse models and the emergence of classicality}

The Ghirardi-Rimini-Weber (GRW) model of spontaneous \wf~collapse as well as
\BM~are the basis of the research program outlined by John S.  Bell in 1987, the
GRW model was proposed as a solution of the measurement problem of quantum
mechanics and involves a stochastic and nonlinear modification of the \se. It
deviates very little from the \se~for microscopic systems but efficiently
suppresses, for macroscopic systems, superpositions of macroscopically different
states and as suggested by Bell, the primitive ontology, or local beables, of
the model are the discrete set of space-time points, at which the collapses are
centered. This set is random with distribution determined by the initial
wavefunction.

The aim of the dynamical reduction program is then modify the \se, by
introducing new terms having the following properties:

\begin{itemize}
    \item They must be \textit{non-linear}, as one wants to break the superposition principle
        at the macroscopic level and assure the localization of the \wf~of
        macro-objects.
    \item They must be \textit{stochastic} because, when describing measurent-like
        situations, is neccesary explain why the outcomes occur ramdonly; more
        than this, is neccesary explain why are they distributed according to
        the Born probability rule
    \item The must be an \textit{amplification mechanism} according to which the new
        terms have negligible effects on the dynamics of microscopic systems,
        but at the same time, their effects becomes very strong for large
        many-particle systems such as macroscopic objects, in order to recover
        their classical-like behavior.
\end{itemize}

We limit the presentation to the GRW model, is neccesary clarify this because
there are other models of spontaneous wave function collapse (collapse models).
In particular, we present the formulation of the GRW model due to J.S. Bell.

Let us consider a system of $N$ particles which for simplicity we take to be
scalar. This system in the GRW model is described by a wave function
$\psi(x_1,x_2,\dots x_N)$ belonging to the Hilbert space $\mathit{L}^2(\Re^{3N})$
and at random times, each particle experiences a sudden jump of the form:

\[
    \psi_t(x_1,x_2,\dots x_N) \to
    \frac{\hat{L}_n(x)\psi_t(x_1,x_2,...x_N)}{\|\hat{L}_n(x)\psi_t(x_1,x_2,...x_N)\|}
\]

Where $\psi_t(x_1,x_2,\dots x_N)$ is the state vector of the whole system at
time $t$, immediately prior to the jump process. $\hat{L}_n(x)$ is a linear
operator called the \textit{reduction operator} (the operator on whose
eigenfunctions one wants to reduce the statevector, as a consequence of the
collapse mechanism) which is conventionally chosen equal to:

\be
    \label{LocOp}
    \hat{L}_n(x) = \sqrt[4]{\left(\frac{\alpha}{\pi}\right)^3}
    e^{-\frac{\alpha}{2} (\hat{q}_n -x)^2}
\ee

Where $\alpha$ is a new parameter of the model with sets that the width of the
localization process, and $\hat{q}_n$ is the possition operator associated to
the $n-$th particle; the random variable $x$ correspond to the place where the
jumps occurs. Between two consecutive jumps, the state vector evolves according
to the \se.

The probability density of a jump taking place at the possition $x$ for the
$n-$th particle is given by:

\[
    p_n(x) \equiv \|\hat{L}_n(x)\psi_t(x_1,x_2,\dots x_N)\|^2
\]

And the probability densities for the different particles are independent.
Finally it is assumed that the jumps are distributed in time like a Poissonian
process with the frecuency $\lambda$, which is the second new parameter of the
model.

The standard numerical values for $\lambda$ and $\alpha$ are:

\[
    \lambda \simeq 10^{-16} s^{-1}, \qquad \alpha \simeq 10^{10} cm^{-2}
\]

Then, the first parameter(the collapse strength) sets the strength of the
collapse mechanics for one particle, and through the \textit{amplification
mechanism} also for a macroscopic system. And $r_C = 1/ \sqrt{\alpha}$ (the
spatial correlation function of the noise, which determines the width of the
localization gaussian in \eqref{LocOp}) tell which kind of superpositions are
effectively supressed, and which not. In the case of a compositive system, an
object, in a superposition of two different states, distant relative to each
other more than $r_C$. The collapse mechanism becomes difficult to ulfold
analytically, due to the complexity of the dynamics for the system. Nevertheless
is possible work out an easy approximative rule to estimate the effect. The
center of mass of the system collapses with a rate

\[
    \Gamma = \lambda n^2 N
\]

where $\lambda$ is the collapse rate of a single constituent (a nucleon), $n$ is
the number of constituents contained in a volume of radius $r_C$ (therefore it
is a measure of the density of the system) and $N$ counts how many such volumes
can be accommodated in the system (therefore being a measure of the total volume
of the system).

Finally, let the $m_n$ be the mass associated to the $n-$th ``particle'' of the system;
then the function:

\[
    \rho_{t}^{(n)}(X_n) \equiv m_n \int d^3x_1 \dots d^3x_{n-1}
    d^2x_{n+1}\dots d^3x_N \: |\psi_t(x_1,x_2,\dots x_N)|^2
\]

represents the \textit{density of mass} of that ``particle'' in space, at time t.

With the help of these axioms the GRW model tries to provide a unified
description of all physical phenomena, at least at the non-relativistic level,
and a consistent solution to the emergence of classicality. This because the
collapse rate of a composite system scales with its size: the bigger the system,
the faster the collapse. This is precisely the reason why collapse models can
accommodate, within a single dynamical equation, both the quantum properties of
microscopic systems (few constituents) and the classical properties of
macroscopic objects (many constituents). In doing this, collapse models explain
the quantum-to-classical transition, how the probabilistic and wavy nature of
atoms and molecules gives rise to the world of classical physics as we
experience it, when they glue together to form larger and larger objects, ie,
the wave functions of macro-objets are almost allways very well localized in
space, so well localized that their centers of mass behave, for all practical
purposes, like point-particles moving according to Newton's laws. In the other
hand at the microscopic level, quantum system behaves almost exactly as
predicted by \oqm, and in measurement-like situation, e.g. of the Von neumann
type reproduces as a consequence of the modified dynamics the Born probability
rule and the collapse of the \wf.

The GRW then leads to a collapse process that is adequate to explain all present
experimental results but, inevitably, although it solves some problems it
produces others. In particular. It has been criticized because in the same way
of all most important dynamical reduction models, which aim at localizing
wavefunction in space, exhibit the typical feature of violating the energy
conservation for isolated systems. In fact if one choose $\hat L$ to be the
position operator $\hat q$ or a function of it, like \eqref{LocOp} (the reason
to use the position operator is that is the most natural candidate for
localizing wavefunctions in space), then it can be shown that the amplification
mechanism induces larger and larger fluctuations in the momentum space as one
can roughly infer using the uncertainty principle: The smaller the uncertainty
in position the bigger uncertainty in momentum. Such fluctuations, in turn,
determine the growth of the energy of the system, which eventually diverges for
large times. Then, apparenly, the energy nonconservation appears to be an
intrinsic property of collapse models since the stochastic process driving the
reduction mechanism seems to be directly responsible for such a violation.

Now, it is neccesary to stress some points about the world-view provides by the
GRW model. According to the ontological interpretation of this theory, there are
no particles at all in the theory! There are only distributions of masses which,
at the microscopic level, are in general quite spread out. An electron, for
example, is not a point following a trajectory as in \BM~ but a wavy system
diffusing in space. When in the double-slit experiment, or more specific in our
case, in the \Ae~, it goes through the both of them, as a classical water-wave
would do. The peculiarity of the electron, which qualifies it as a quantum
system, is that when we try to localize it in space by letting it interacting
with a measuring devise, e.g. a photographic plate, then, according to the
localization process and because the interaction with the plate, its wave
function very rapidly shrinks in space till is gets localized to a spot, the
spot where the plate is impressed and which represents the outcome of the
measurement. The big difference between GRW and \oqm~is that such a behaviour is
not accounted by a crude ``instantaneous collapse'' prescription, with has no
explanation within the theory, conversely it is a direct consequence of the
universal dynamics of the GRW model, also in it framework even though the GRW
model contains no particles at all, is usually referred to microsystems as
``particles'', just for a matter of convenience, this because also macroscopic
objects in this model are waves, their centers of mass are not mathematical
points, rather they are represented by some function defined throughout space,
but with the property of be always almost perfectly located in space, which
means that the wave function associated to their centers of mass are appreciably
different from zero only within a very tiny region of space, so tiny that can be
consider point-like for all practical purposes such dynamics is governed by
Newton's laws.

In \cite{Toros16} it is shown how Bohmian trajectories become classical when
particles glue together to form macroscopic objects, which unavoidably interact
with the surrounding environment, the dynamics of this process obeys in a first
approximation a collapse process similar to the postulated in the GRW model.
This might open the way to finding an underlying theory. These two theories are
always presented as dichotomical, but as shown in \cite{Allori02} they have much
more in common that one would expect at first sight. The differences are less
profound than the similarities and then using the common structure of them would
be possible formulate a new theory with benefits from \BM~and the GRW model and
(collapse models in general).

\subsection{The non-local ontology of the Ghirardi-Rimini-Weber model}

\bqu{
``Either the \wf~as given by the \se, is not everything, or it is not right.''
Bell
}\equ

How we as shown in this work Bohmian trajectories serve to explain quantum
mechanics in terms of a materialistic theory, providing the foundation for the
claim that a fundamental physics focused on an objective reality outside the
mind is possible. But this is not the case of the GRW model where only the wave
function is regarded as existing, the basic problem of this theory is that never
talks about matter, and then would not form an adequate description of the
world. Strictly speaking, in a world governed by such a theory there exist no
matter neither a clear link between the objective reality and the mathematical
variables of the theory.

Even worse, if we use the GRW model to deal with the \Ae~paradox we get the same
problem of \oqm. In general the problem that it remains to be clarified is in
what the way the \wf s $\psi_{\Sigma} \wedge \psi_{\Sigma'}$ associated with
different regions $\Sigma \wedge \Sigma'$ of the space have to be compatible to
be regarded as describing a consistent reality. In the case $\Sigma \wedge
\Sigma'$ have a large portion in common like in the \Ae, is necessary that
$\psi_{\Sigma} \wedge \psi_{\Sigma'}$ describe the same reality on $\Sigma \cap
\Sigma'$. This is one of the problems in the GRW model and we need define
clearly what that reality is and how the \wf~influences that reality.

Assuming that $\psi_{\Sigma} \wedge \psi_{\Sigma'}$, are the \wf s associated to
different regions of space, any of them, suppose \wf~$\psi_{\Sigma}$ can be
constructed by superposing, in the same region of space, waves of slightly
different wavelengths, but with phases and amplitudes chosen to make the
superposition constructive in the desired region and destructive outside it.
Mathematically, we can carry out this superposition by means of non-local Fourier
analysis. For simplicity, we are going to consider a one-dimensional wave
packet. Then, we construct the packet $\psi_{\Sigma}(x,t)$ by superposing plane
waves(propagating along the $x$-axis) of different frecuencies (or wavelengths):

\be
\label{wpack}
\psi_{\Sigma}(x,t) = \frac{1}{\sqrt{2\pi}} \int_{-\infty}^{+\infty} \phi(k) e^{i(kx-\omega
t)} dk
\ee

where $\phi(k)$ is the amplitude of the wave packet. The packet is intended to
describe a particle confined to a one-dimensional region, in this case a
particle moving along the $x$-axis. We want to look at form of the packet at a
given time. Choosing this time to be $t = 0$ and abbreviating
$\psi_{\Sigma}(x,0)$ by $\psi_{\Sigma_0}$, we can reduce \eqref{wpack} to:

\be
\label{iFtr}
\psi_{\Sigma_0}(x) = \frac{1}{\sqrt{2\pi}} \int_{-\infty}^{+\infty} \phi(k)
e^{ikx} dk
\ee

and

\be
\label{Ftr}
\phi(k) = \frac{1}{\sqrt{2\pi}} \int_{-\infty}^{+\infty} \psi_{\Sigma_0}(x)
e^{-ikx} dk
\ee

where $\phi_k$ is the Fourier transform of $\psi_{\Sigma_0}(x)$, and where

\be
\label{Fkernel}
e^{ikx} = cos(kx) + isen(kx)
\ee

is the kernel of the transformation.

We refer to \eqref{iFtr} as the synthesis equation and to \eqref{Ftr} as the
analysis equation for $\psi_{\Sigma_0}$. Then, is expected that the packet whose
form is determined in this way by the $x$-dependence of $\psi_{\Sigma_0}$, does
indeed have the required property of localization: peaks at $x = 0$ and vanishes
far away from $x = 0$. The idea is that when $x \to 0$ then $e^{ikx} \to 1$;
because the waves of different frequencies interfere constructively, and far
away form $x = 0$ (i.e., $\|x\| \gg 0$) the phase \eqref{Fkernel} goes through many
periods leading to violent oscillations, thereby yielding destructive
interference.

However, what we actually are doing here is taking the function that represents
the information available on the physical situation, the \wf. The next step is
to make the analysis using \eqref{Ftr}, that is, the decomposition in terms of
the kernel \eqref{Fkernel} (i.e., the trigonometric functions sine and cosine
that are defined from minus infinity $(-\infty)$ to infinity $(\infty)$ both in
space and time) which amounts to find the coefficients for the sines and cosines
or, in the case of a non-periodic function, the coefficient function $\phi_k$.
After that, once these coefficients are known, one normally makes the synthesis
of the initial function using \eqref{iFtr}.

Thus, when we sum a certain number of these waves to synthesise $\psi_{\Sigma}$
a problem arise. The \wf~$\psi_{\Sigma}$ that according to the GRW model
contains all the information of the system and is used to obtain the \textit{the
density of mass in space} is being synthesized with Fourier analysis and
this analysis is non-local!, i.e., it kernel, it's building blocks, are harmonic
plane waves, and these harmonic plane waves are no more than sine and cosine
functions, which are defined in whole space and in hole time. Then, at the
microscopic level, the `particles' according to the GRW model are in general
quite spread out in a wavy system diffusing in space.

This is a serious problem, because within the GRW model with non-local Fourier
analysis, all the `particles' of the universe compose an infinite sea of
\textit{density of mass}, with only a potential existence that the collapse
process materializes at a given position, and in this sense the GRW model fall
in the same idealistic approach that \oqm.

\subsection{Perspectives: A theory of exclusively Local Beables}

Using non-local Fourier analysis, as seen before, we were able to obtain the
\textit{density of mass in space} in the GRW model. However, this analysis is a
non-local one, that is, it core are unlimited waves, either in space and time.
Then, the non-local ontology of the GRW model comes from the way we
mathematically represent the \wf.

We now propose a second possibility, This possibility is use Wavelet analysis.
Wavelets were invented to overcome the shortcomings of Fourier analysis. With
wavelets we can synthesize the \wf~without gathering information contained in all
space and time. Both analysis allows us to reconstruct the \wf.  However, while
Fourier analysis uses mathematical entities existing in whole space and whole
time, Wavelet analysis use mathematical entities existing only locally.

For many decades scientist have wanted more appropriate functions than the sines
and cosines, which are the basis of local Fourier analysis. These functions are
non-local, they therefore do a very poor job in approximating sharp spikes. But
with wavelet analysis, we can use approximating functions that are contained
neatly in finite domains.

In 1975 Bell introduced the term beable to name whatever is posited, by a
candidate theory, as corresponding directly to something that is physically
real, i.e., independent of any observation. He then divides beables into two
categories, local and non-local. In this sense, a theory as proposed above can
be consider a theory of exclusive local beables. And it would provide also a
concrete example of an empirical viable quantum theory in whose formulation the
\wf~on configuration space does not appear, i.e., it is a theory according to
which nothing corresponding to the configuration space \wf~need actually exist.

The idea with this is it might be possible to construct a plausible,
empirically, viable theory of this kind, and to recommend this as an interesting
and perhaps-fruitful program for future research. The theory of exclusive local
beables put forward here is only an intended of toy model, to illustrate in
principle that this kind of theory can, after all, be constructed. The challenge
is then, for a system of $N$ particles moving in three spacial dimension, the
theoretical description would be of $N$ \wf s. Now, there are some things we
have to consider:

\begin{itemize}

\item The \se~for an $N$-particle system is not a set of N (interacting) waves, each
propagating in 3-space. It is a single wave propagating in $3N$-dimensional
\textit{configuration space} for the system.

\item Schr\"odinger \wf~for such an $N$-particle system living in the
    \textit{configuration space} is by this characteristic a non-local agent,
    and any theory which aims to a correct description of nature must be
    non-local.

\end{itemize}

Fractal geometry is based on the idea of self-similar forms. To be self similar,
a shape must be able to be divided into parts that are smaller copies which are
more or less similar to the whole. Because of the smaller similar divisions of
fractals, they appear similar at all magnifications. In this sense, non-locality
could be recover in this model conserving the individuality of the beables
through the use of multiresolution-based or fractal wavelets (Multi-resolution
analysis (MRA)). A local beable of this nature is not a particle lying in
space-time but, conversely, all the space-time is included in each beable.

In \BM~the quantum potential leads to the notion of an ``unbroken wholeness''
and is the responsible for the non-local causality. On the other hand scale
relativity with Hausdorff dimension $2$ is intimately related to the \se~and
quantum mechanics. If Einstein showed that space-time was curved, scale
relativity shows that it is not only curved, but also fractal, i.e., that a
space which is continuous and non-differentiable is necessarily fractal. It
means that such a space depends on scale.

Importantly, scale relativity does not merely describe fractal objects in a given
space. Instead, it is space itself which is fractal. To understand what a
fractal space means requires to study not just fractal curves, but also fractal
surfaces, fractal volumes, etc.

Mathematically, a fractal space-time is defined as a nondifferentiable
generalization of Riemannian geometry. Such a fractal space-time geometry is the
natural choice to develop this new principle of relativity, in the same way that
curved geometries were needed to develop Einstein's theory of general
relativity.

Scale relativity predicts the existence of trajectories varying according to
scale transformations. Those trajectories are not rectifiable, i.e., they are
fractal. The apparently disordered motions of particles at the quantum scale are
described in terms of motions following fractal geodesics.

As has been shown in \cite{Carroll12}, a recent paper of Kobelev describes a
Leibniz type of fractional derivative and one can relate fractional calculus
with fractal structures, then is shown that if is possible write a meaningful
\se~with Kobolev derivatives ($\alpha$-derivatives) then there will be a
corresponding fractional quantum potential.

MRA could be a useful method for the construction of the neccesary nonlinear
bases, since the linear harmonic analysis is inadequate for describing nonlinear
systems. The wavelets in MRA are functions that have a space dependent scale
which renders them an invaluable tool for analyzing multi-scale phenomena.

\section{Conclusion}
The aim of this thesis was use the \Ae~`paradox' to try to answer the question,
{\it with what does the quantum mechanics actually deal ?}. The answer given by
Bohr (and within the framework of the usual interpretation the only possible
consistent answer) is that it deals not with the properties of the objective
reality as such but, rather, with nothing more than the relationships among the
observable large-scale phenomena. The general point of view of Bohr has been
given its most consistent and systematic expression, in terms of the \PC.
The \PC~states that we are restricted to complementary pairs of inherently
imprecisely defined concepts, such as position and momentum, wave and particle,
etc. The maximum degree of precision of either member of such a pair is
reciprocally related to that of the opposite member. This assumption leads to
the conclusion that the basic properties of matter can never be understood
rationally, e.g., an atom has no properties at all when it is not observed.
Indeed, one may say that its only mode of being is to be observed; for the
notion of an atom existing with uniquely definable properties of its own even
when it is not interacting with a piece of observing apparatus, is meaningless
within the framework of this point of view.
We have shown however through the analysis of the \Ae~`paradox' an example of
why this response is  unsatisfactory and leads to the current crisis in quantum
physics, we have shown that the origin of this crisis consists in the rejection
of an objective reality existing outside the mind, i.e., by try justify the
tenets of quantum philosophy, namely `philosophical idealism', through quantum
physics, limiting the possible develop of quantum physics with the limits
implicit in the \PC. We have shown that other answer to the question is possible
using \BM. What does \BM~contribute here? In a word, everything! \BM~is a
beautiful example of a theory that is satisfactory as a fundamental physical
theory, a fundamental physical theory very different from classical mechanics,
but we that it be as clear as classical mechanics.
\BM~is a counterexample to all claims that a rational account of quantum
phenomena is impossible. It is also a counterexample to the claim that
\qm~proves that nature is intrinsically random that there is no way that
determinism can ever be reinstated in the fundamental description of nature.
This work is a simple example of why physics must be about the objective reality
and need recognizes instead that a quantum theory must describe such a reality.

\bibliographystyle{acm}
\bibliography{Bibliography}
\end{document}